\documentclass[a4]{iopart}  
\usepackage[T1]{fontenc}
\usepackage[latin1]{inputenc}  
\usepackage{hyperref}  
\usepackage{geometry}
\usepackage{graphicx}  
\usepackage{amssymb}  
    
\date{\today}  
\def \be{\begin{equation}}
\def \bea{\begin{eqnarray}}
\def \eea{\end{eqnarray}}
\def \ee{\end{equation}}
\def \no{\nonumber}

\def \Ldot{\dot L}

\def \h{\frac{1}{2}}
\def \Q{{\cal Q}}
\def \M{{\cal M}}
\def \U{{\cal U}}
\def \K{{\cal K}}
\def \Kbar {{\bar \K}}
\def\lsim{\mathrel{\rlap{\lower4pt\hbox{\hskip1pt$\sim$}}
    \raise1pt\hbox{$<$}}}                
\def\gsim{\mathrel{\rlap{\lower4pt\hbox{\hskip1pt$\sim$}}
    \raise1pt\hbox{$>$}}}                

\begin{document}  
\title{Time Delay Interferometry for LISA with one arm dysfunctional}  
\author{S. V. Dhurandhar$^1$, K. Rajesh Nayak$^2$ and J-Y. Vinet$^3$}  
\address{ $^1$ IUCAA, Postbag 4, Ganeshkind, Pune - 411 007, India.     
\\$^2$ IISER-Kolkata, PO: BCKV Campus Main Office, Mohanpur - 741252, India. 
\\ $^3$ ARTEMIS, Observatoire de la Cote d'Azur, BP 4229, 06304 Nice, France.
}  
  
\begin{abstract}  

In order to attain the requisite sensitivity for LISA - a joint space mission of the ESA and NASA- the laser frequency noise must be suppressed below the secondary noises such as the optical path noise, acceleration noise etc. By combining six appropriately time-delayed data streams containing fractional Doppler shifts - a technique called time delay interferometry (TDI) - the laser frequency noise may be adequately suppressed. 
We consider the general model of LISA where the armlengths vary with time, so that second generation TDI are relevant. However, we must envisage the possibility, that not all the optical links of LISA will be operating at all times, and therefore, we here consider the case of LISA operating with two arms only. As shown earlier in the literature, obtaining even approximate solutions of TDI to the general problem is very difficult. Since here only four optical links are relevant, the algebraic problem simplifies considerably. We are then able to exhibit a large number of solutions (from mathematical point of view an infinite number) and further present an algorithm to generate these solutions. 

\end{abstract} 


\maketitle  
\section{Introduction \label{SC:1}}  

LISA - Laser Interferometric Space  Antenna - is a proposed ESA-NASA mission which will use coherent laser beams exchanged between three identical spacecraft forming a giant (almost) equilateral triangle of side $5 \times 10^6$ kilometres to observe and detect low frequency cosmic GW \cite{ESANASA}. 
\par
Laser frequency noise dominates the other secondary noises, such as optical path noise, acceleration noise by 7 or 8 orders of magnitude, and must be removed if LISA is to achieve the required sensitivity of $h \sim 10^{-22}$, where $h$ is the metric perturbation caused by a gravitational wave. In LISA, six data streams arise from the exchange of laser beams between the three spacecraft  approximately 5 million km apart. These six streams produce redundancy in the data which can be used to suppress the laser frequency noise by the technique called time-delay interferometry (TDI) in which  the six data streams are combined with appropriate time-delays \cite{ETA}. A mathematical foundation for the TDI problem for the static LISA was given in \cite{DNV}, where it was shown that the data combinations cancelling laser frequency noise formed the {\it module of syzygies} over the polynomial ring of time-delay operators. For the static LISA, the polynomial ring was in three variables and also commutative. The generators of the module were obtained via Gr\"obner basis methods. This scheme can be extended in a straight forward way to include the (Sagnac) effect arising from the rotation of LISA, where the up-down optical links are unequal and as a consequence, now one has six optical links, but the armlengths are still constant in time. These are the modified Ist generation TDI. Here the polynomial ring is still commutative although over six indeterminates. The generators of the module were obtained in \cite{NV}. In the general case of time varying armlengths, the polynomial ring is in six variables (the six optical links) but now it is noncommutative. The algebraic problem in the general case has been discussed in \cite{SD}, but the solution to the general problem seems extremely difficult. We still have a linear system which leads to a module (a left module), but it seems difficult to obtain its generators in general. In the non-commutative case, even the Gr\"obner basis algorithm may not terminate. 
\par
We must envisage the possibility that not all optical links of LISA will be operating at all times for various reasons  like technical failure for instance or even the operating costs. Therefore, it is important to discuss the question when not all the links operate. Here we look at a specific situation where the data is available from only two arms or four optical links. This should not much affect the information that can be extracted from the data, because this is essentially a Michelson configuration which is known to be quite useful. The practical advantage for this case is that the algebraic problem simplifies considerably and therefore becomes tractable. We can reduce the problem to that of only one linear constraint on two polynomials, although the equation is still noncommutative. We show that we can generate an infinity of solutions by using a combinatorial algebraic approach which lists all such solutions in a systematic way. The solutions - that is the laser frequency noise is suppressed for these time-delayed  combinations - are approximate in the sense that the ${\ddot L}$ and $\Ldot^2$ terms are ignored in the calculation, where $L(t)$ is the generic length of the LISA's arm or the optical link and the 'dot' denotes derivative with respect to time. The solutions are based on vanishing commutators (in this approximation). We enumerate such commutators and for each such commutator there is a corresponding solution. We present an algorithm to construct these solutions.  
\par
A geometric combinatorial approach was adopted in \cite{Vallis} where several solutions were exhibited. Our approach is algebraic where several operations are algebraic operations on strings of operators. The algebraic approach has the advantage of easy manipulation of data streams, although some geometrical insight could be at a premium. The analysis presented here can be used also for other future space detectors like ASTROD where again a similar situation arises \cite{WTN}.  

\section{Preliminaries and notation}

\begin{figure}[h!]  
\centering  
\includegraphics[width = 0.4\textwidth]{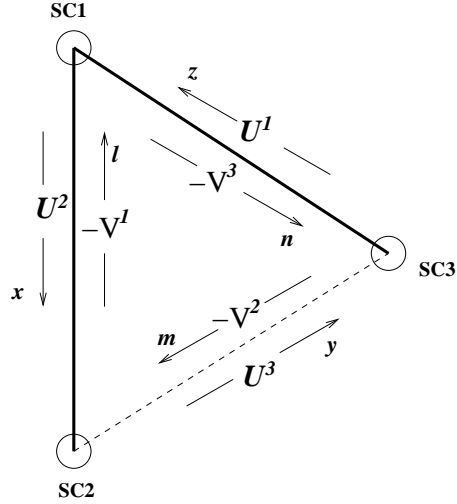}  
\caption{The beams and the corresponding time-delays are shown schematically in the figure. The functional  arms are depicted with a continuous line while the dysfunctional arm is shown with a dashed line.}  
\label{schematic}  
\end{figure}

In this section we recapitulate the results obtained in the earlier literature already mentioned and also set up notation. In the literature on TDI of LISA there are several notations - here we choose the one which seems simplest for our purpose. We follow the notation and conventions of \cite{DNV} and \cite{NV}. The six links are denoted by $U^i, V^i , i = 1, 2, 3$. The time-delay operator for the link $U^2$ from S/C 1 to S/C 2 or $1 \longrightarrow 2$ is denoted by $x$ in \cite{NV} and so on in a cyclic fashion. The delay operators in the other sense are denoted by $l, m, n$;  the link $-V^1$ from $2 \longrightarrow 1$ by $l$ and similarly the links $V^2, V^3$ are defined through cyclic permutation.   Figure \ref{schematic} depicts the optical links as described. 
\par
Let $C_i (t) = \Delta \nu_i (t)/\nu_0$ represent the laser frequency noise in S/C $i$. Let $j$ be the delay operator corresponding to the variable armlength $L_j (t)$, i.e. $ j C_i (t) = C_i (t - L_j (t))$. Then we have, 
\bea
U^1 = C_1 - z C_3 \,, \no \\
V^1 = l C_2 - C_1 \,.
\eea
The other links in terms of $C_i (t)$ are obtained by cyclic permutations. Also in the $U^i, V^i$ we have not included contributions from the secondary noises, gravitational wave signal etc. since here our aim is to deal with laser frequency noise only. Any observable $X$ is written as: 
\be
X = p_i V^i + q_i U^i \,,
\ee
where $p_i, q_i, i = 1,2,3$ are polynomials in the variables $x,y,z,l,m,n$. Thus $X$ is specified by giving the six tuple polynomial vector $(p_i, q_i)$.  Writing out the $(V^i, U^i)$ in terms of the laser noises $C_i (t)$, and in order that the laser frequency noise cancel for arbitrary functions $C_i (t)$,  the polynomials $(p_i, q_i)$ must satisfy the equations:
\bea
p_1 - q_1 +  q_2 x -  p_3 n &=& 0, \no \\
p_2 - q_2 +  q_3 y -  p_1  l &=& 0, \no \\
p_3 - q_3 +  q_1 z -  p_2 m &=& 0 \,,
\label{lneq}
\eea 
where one must remember that the order of the operators is important when the armlengths are time-varying. The solutions (if they exist) to Eq. (\ref{lneq}) are the second generation TDI. One finds that at least in the approximation we consider they do exist. For a reasonably optimised model of LISA, say given in \cite{DNV08}, $\ddot L \sim 10^{-6}$ metres/sec$^2$ and thus even if one considers say 20 successive optical paths, that is, about $\Delta t \sim 330$ seconds of light travel time, $\Delta t^2 {\ddot L} \sim 0.1$ metre. This is well below few metres and thus can be neglected in the residual laser noise computation. Moreover, ${\dot L}^2$ terms (and higher order) can be dropped since they are of the order of $\lsim 10^{-15}$ (they come with a factor $1/c^2$) which is much smaller than 1 part in $10^8$, which is the level at which the laser frequency noise must be cancelled. Thus, we keep terms only to the first degree in $\Ldot$ and also neglect higher time derivative terms in $L$. 
\par
The solutions to these equations are the polynomial vectors $X = (p_i, q_i)$ and they form a left module $\M$ over the non-commutative ring $\K = \Q(x, y, z, l, m, n)$, where $\Q$ is the field of rational numbers and $x, y, z, l, m, n$ are the indeterminates. A left module means that one can multiply a solution $(p_i, q_i)$ from the {\it left} by any polynomial in $\K$, then it is also a solution to the equations (\ref{lneq}) and therefore in the module $\M$. It is difficult to solve these equations in general for the non-commutative case. Previously we have shown in \cite{SD} that the ring $\K$ can be reduced to the quotient ring $\Kbar$ which is defined by quotienting $\K$ by the ideal $\U$ generated by vanishing commutators. Consider commutators of the form $[x_1 x_2 ... x_n, y_1 y_2 ... y_n]$ where $n \geq 2$ and $x_k$ or $y_m$ represents any one of the delay operators  $x, y, z, l, m, n$. Upto the order of approximation we work in:
\be
[x_1 x_2 ... x_n, y_1 y_2 ... y_n] = \sum_{k = 1}^n L_{x_k} \sum_{m = 1}^n \Ldot_{y_m} - \sum_{m = 1}^n L_{y_m} \sum_{k = 1}^n \Ldot_{x_k} \,. 
\ee
From this equation it immediately follows that if the set of operators $y_1, y_2, ..., y_n$ is a permutation of the operators $x_1, x_2, ..., x_n$, then the commutator:
\be
[x_1 x_2 ... x_n, y_1 y_2 ... y_n] = 0 \,,
\label{xct_cmtr}
\ee
upto the order of the approximation. The ideal $\U$ is generated by such commutators. 
\par
Although we may be able to apply non-commutative Gr\"obner basis methods, the general solution seems quite difficult. However, we could consider a simpler problem of one arm of LISA being nonfunctional. Then we have a simpler situation which turns out to be tractable.

\section{Solutions with one arm nonfunctional}

We arbitrarily choose the non-functional arm to be the one connecting S/C 2 and S/C 3. This means from our labelling that the polynomials are now restricted to only the four variables $x, l, z, n$ and we can put the polynomials $p_2 = q_3 = 0$. This simplifies the equations (\ref{lneq}) considerably. The second and third of the equations in (\ref{lneq}) simplify to:
\bea 
q_2 &=& - p_1 l \,, \no \\
p_3 &=& - q_1 z \,.
\label{eqsimp}
\eea
Substituting these in the first equation, we need to solve just one equation:
\be
p_1 (1 - lx) - q_1 (1 - zn) = 0 \,,
\ee
to obtain the full polynomial vector as a solution. It is clear that solutions are of the Michelson type denoted by $X$. One such solution has already been given in the literature \cite{ETA,Vallis}. This solution in our notation is:
\bea
p_1 &=& 1 - zn - znlx + lx (zn)^2 \, \no \\
q_1 &=& 1 - lx - lxzn + zn (lx)^2 \,.
\label{X1}
\eea  
Writing,
\be
\Delta = p_1 (1 - lx) - q_1 (1 - zn) \,,
\ee
we get for (\ref{X1}), $\Delta = [znlx, lxzn]$ which is a commutator that vanishes. Thus it is an element of $\U$ and (\ref{X1}) is a solution (over the quotient ring). This is already well known.
\par
{\it What we would like to emphasise is that there are more solutions of this type - in fact there are infinite number of such solutions.} In what follows we give a systematic way to enumerate and list such solutions. 
\par
Notice that the operators $lx$ and $zn$ occur together. Physically, these operators correspond to round trips. We denote these round trip operators by $a = lx$ and $b = zn$, then we can write the previous equation and solution respectively, in a compact way as:
\be
p_1 (1 - a) - q_1 (1 - b) = 0 \,,
\label{4lnk}
\ee
and,
\bea
p_1 &=& 1 - b - ba + ab^2 \,, \no \\
q_1 &=& 1 - a - ab + ba^2 \,. 
\label{X1'}
\eea
We can substitute this in $\Delta = p_1 (1 - a) - q_1 (1 - b)$ to obtain $\Delta = [ba, ab]$ which vanishes and hence (\ref{X1'}) is a solution.
\par
But we can proceed further and write:
\bea
p_1' &=& p_1 - b a^2 b \,, \no \\
q_1' &=& q_1 - a b^2 a \,,
\eea
to obtain:
\be
\Delta' = b a^2 ba - a b^2 ab \,.
\ee
$\Delta'$ however does not vanish - it is also not a commutator - and hence $p_1', q_1'$ is not a solution. Continuing further with,
\bea
p_1'' &=& p_1' + a b^2 a b  \,, \no \\
q_1'' &=& q_1' + b a^2 b a\,,
\eea
we get, $\Delta'' = b a^2 bab - a b^2 aba$ which also does not vanish, again not a commutator.
\par
This process can be continued in such a way so that lower degree terms cancel and so as to get another vanishing commutator but of higher degree. The next commutator is obtained by adding two more terms of successively higher degree in a similar way. The result is:
\bea
p_1 &=& 1 - b - ba + ab^2 - b a^2 b + a b^2 ab + ab^2 aba - ba^2 bab^2 \,, \no \\
q_1 &=& 1 - a - ab + ba^2 - a b^2 a + b a^2 ba + ba^2 bab - ab^2 aba^2 \,.
\label{X2}
\eea
Note that $q_1$ is obtained from $p_1$ by interchanging $a$ and $b$ and vice versa. This is a symmetry. The expressions for $p_1$ and $q_1$ in Eq. (\ref{X2}) give the vanishing commutator:
\be
\Delta = [a b^2 a, b a^2 b] \,,
\ee
and hence (\ref{X2}) is a solution.
\par
There are two more solutions with eight terms. One of them is:
\bea
p_1 &=& 1 + a - b^2 - b^2 a - b^2 a^2 - b^2 a^3 + a^2 b^4 + a^2 b^4 a \no \\
    &=& (1 - b^2 - b^2 a^2 + a^2 b^4)(1 + a) \,, \no \\
q_1 &=& 1 + b - a^2 - a^2 b - a^2 b^2 - a^2 b^3 + b^2 a^4 + b^2 a^4 b \no \\ 
    &=& (1 - a^2 - a^2 b^2 + b^2 a^4)(1 + b) \,,
\eea
with $\Delta = [b^2 a^2, a^2 b^2]$ which is also a vanishing commutator. This solution looks something like the square of the solution (\ref{X1'}) obtained by Tinto et al (see \cite{ETA}) multiplied by linear factors in order ensure successive term cancellation. Here one starts with $p_1 = 1 + a ...$ and continues until one reaches the above commutator. 
\par
The third solution corresponds to the commutator $[baba, abab]$. This solution is given by:
\bea
p_1 &=& 1 - b + ab - bab - baba + abab^2 - baba^2 b + abab^2 ab \,, \no \\
q_1 &=& 1 - a + ba - aba - abab + baba^2 - abab^2 a + baba^2 ba \,.
\eea
\par
The following observations are in order: the solutions correspond to the vanishing commutators as follows; there is only one commutator at degree 4, namely, $[ba, ab]$; of degree 8, there are three commutators: 
$[a^2 b^2, b^2 a^2],~ [abab, baba],~ [a b^2 a, b a^2 b]$. And these are the only three commutators at degree 8.
Also there are equal number of $a$'s and $b$'s in each string of the commutators. To each of the vanishing commutators, there is a corresponding polynomial vector solution $(p_1, q_1)$. The rest of the solution vector is obtained from Eq.(\ref{eqsimp}) which then yields $p_3$ and $q_2$.  
\par 
These are solutions only upto degree 7. Clearly one can construct solutions of higher degree. We do this in a combinatorial way in the next section. We enumerate the solutions in a given order - called the lexicographic order - and also present an algorithm to obtain the polynomials $p_1$ and $q_1$.
 
\section{Solutions of higher degree}

\subsection{Listing the commutators}
 
In order to proceed further, let us first define an interchange operator $I$ between the symbols $a$ and $b$, that is, given a string $s$ composed of $a$ and $b$, the string $t = Is$ is obtained by replacing $a$ by $b$ in $s$. A string is a polynomial with only one term usually also called a monomial in the literature. The length of a string is the degree of the monomial. We will use these terms interchangably, but that should not lead to any confusion. Note that all the vanishing commutators which have appeared corresponding to a solution have the form $[s, Is]$. Also since the string $Is$ must be a permutation of the string $s$, in order that the commutator $[s, Is]$ vanish, there must be an equal number of $a$'s and $b$'s in the string $s$ of the vanishing commutator. So if we have $n$ number of $a$'s in the string $s$, it must also have the same $n$ number of 
$b$'s, so that the total string $s$ has length $2n$. The commutator is then made up of strings of length $4n$. Note that the operator $I$ can be extended to a polynomial containing strings in an obvious way by defining:
\be
I (s_1 + s_2 + ...) = I s_1 + I s_2 + ... \,.
\ee
Then we easily check that in the foregoing $q_1 = I p_1$, where $(p_1, q_1)$ constitute a solution to 
Eq. (\ref{4lnk}). Also $I^2 = 1$ the identity mapping. $I$ can be considered to be an operator over the polynomial ring over the indeterminates $a$ and $b$.   
\par
In the foregoing we have then considered cases for $n = 1, 2$. For $n = 1$ there is one solution, while for $n = 2$ there are three distinct solutions. In order to proceed further, we must first list the commutators at each $n$. In order to carry out this programme, we must introduce an order among the strings of length $2n$. We choose this order to be {\it lexicographic} or if we consider strings of all lengths, then {\it length lexicographic} \cite{BK}. But since at a time we deal only with strings of a given length $2n$, the lexicographic order suffices. We arbitrarily choose $a < b$. Then at length 2, $aa < ab < ba < bb$. At length 3, $aaa < aab < aba < abb < baa < bab < bba < bbb$ and so on. More generally, for strings of the same length, $s < t$ if at the first position from the left the strings differ, the symbol in string $s$ is less than the symbol in string $t$. In this way we may list all strings in the symbols $a, b$ of a given length. This is all we require. 
\par
Thus at length $2n = 2$ we have just one commutator $[ab, ba]$. At $2n = 4$, we have $[aabb, bbaa], [abab, baba], [abba, baab]$, that is, three commutators; we have also three solutions corresponding to each commutator. We have also taken care to write the first string in the commutator to be less than the second string in the lexicigraphic order chosen. Note that we could list the commutator just by the first string, because the second string in the commutator is obtained by applying the operator $I$ to the first string. We will give an algorithm to obtain a solution, that is, the polynomials $p_1$ and $q_1$ in a unique way given the commutator. Thus listing all the commutators of a string of length $2n$ amounts to listing the solutions $(p_1, q_1)$ having that commutator. The polynomials $p_1$ and $q_1$ have $2n$ terms and degree $2n - 1$. We are now ready to list the commutators $[s, Is]$ with first string $s$ of length $2n$.    
\par
Consider $n = 3$. The string $s$ has length 6 with 3 $a$'s and 3 $b$'s. Listing the strings in lexicographical order we have: $aaabbb < aababb < aabbab < aabbba < abaabb < ababab < ababba < abbaab < abbaba < abbbaa$. We stop here because, the next string is $baaabb = I (abbbaa)$ and $abbbaa$ has occurred in the listing before and hence would lead to minus the commutator and therefore not an independent commutator. This follows from the symmetry encoded in the interchange operator $I$. Thus at $n = 3$ there are 10 independent commutators. Proceeding in this way we find that at $n = 4$ there are 35 such commutators. The general formula for the number of commutators is $^{2n -1} C_{n - 1}$. This formula follows from the following argument. Consider a string of length $2n$, with $n$ number of $a$'s and $b$'s. Because of symmetry between $a$ and $b$, the first 
$a$ is fixed in the string; as seen, the string starting with $b$ does not lead to a new commutator. Thus we must choose $n - 1$ combinations of $a$'s among $2n - 1$ available positions in the remaining string. This is the number of independent commutators which is then just $^{2n -1} C_{n - 1}$.  

\subsection{Algorithm for the solution given the commutator}

The algorithm is obtained by observing the way we got the solutions of degree 7 for $n = 2$. We started with the lowest degree and proceeded towards obtaining the commutator by cancelling terms at successively increasing degree. But now we have been assigned a commutator before hand and we need to find the solutions $(p_1, q_1)$. We therefore need to go in the reverse order. We freely make use of the operator $I$ in this algorithm.
\par
Consider the commutator $\Delta = [s_{2n}, I s_{2n}] = s_{2n} I (s_{2n}) - I (s_{2n}) s_{2n}$ where $s_{2n}$ is a string containing $n$ number of $a$'s and an equal number of $b$'s. 
\begin{enumerate}

\item Consider the first string in the commutator $\Delta$ which we denote by $t_{4n} = s_{2n} I (s_{2n})$. The subscript denotes the length of the string or the degree of the monomial. Now, either $t_{4n}$ ends in $a$ or $b$; that is, $t_{4n} = s_{4n - 1} a$ or $t_{4n} = s_{4n - 1} b$. 

\begin{itemize}
\item If $t_{4n} = s_{4n - 1} a$, then write $t_{4n - 1} = s_{4n - 1}$ or; 
\item if $t_{4n} = s_{4n - 1} b$, then write $t_{4n - 1} = - I s_{4n - 1}$.
\end{itemize}

\item Replace $t_{4n}$ by $t_{4n-1}$ and repeat. That is either we have $t_{4n - 1} = s_{4n - 2} a$ or 
$t_{4n - 1} = s_{4n - 2} b$. If $t_{4n - 1} = s_{4n - 2} a$, then write $t_{4n - 2} = s_{4n - 2}$ otherwise write $t_{4n - 2} = - I s_{4n - 2}$.

\item Continue this process till one arrives at a string of length 0, namely, $t_0$. Note that $t_0 = \pm 1$.

\item Then $p_1 = \sum_{k = 0}^{4n - 1} t_k$ and $q_1 = I p_1$.

\end{enumerate} 

Note that $t_k$ are strings of length $k$. $p_1$ and $q_1$ are polynomials of degree $4n - 1$. 
\par
To illustrate the algorithm explicitly, let us take the case for $n = 3$ and choose one commutator among the 10 possible ones. Let us randomly choose the string $ababba$, so that the commutator is $[ababba, babaab]$. So we have $s_6 = ababba$. Thus $t_{12} = abab^2ababa^2b$. Now $t_{12}$ ends in $b$. Thus we write $t_{12} = s_{11} b$, where $s_{11} = abab^2ababa^2$. Thus choosing the second option in step (ii), we have 
$t_{11} = - I s_{11} = - baba^2babab^2$. This is the highest degree term in the polynomial $p_1$ and for 
$n = 3$, it is of degree 11. The next term is obtained from $t_{11}$. Again $t_{11}$ ends in $b$. Following the steps of the algorithm, $t_{11} = s_{10} b$ and hence $t_{10} = - I s_{10} = abab^2ababa$. Since, $t_{10}$ ends in $a$, $t_9 = s_9 = abab^2abab$. We then have $t_8 = -baba^2bab$ and so on. We must carry out the steps till we reach degree 0. This terminates the algorithm. In this way, we can easily construct the 11th degree polynomial $p_1$ by adding up all the strings $t_k, ~k = 1, 2, ..., 11$. Collecting all the terms and writing out the polynomial $p_1$ explicitly: 
\bea
p_1 &=& 1 - b + ab - bab - baba + abab^2 - baba^2b + abab^2ab \no \\
    &-& baba^2bab + abab^2abab + abab^2ababa - baba^2babab^2 \,.
\eea
 
The polynomial $q_1$ is obtained from $p_1$ by just interchanging $a$ and $b$, that is, $q_1 = I p_1$. The other polynomials  $q_2$ and $p_3$ are obtained from Eq. (\ref{eqsimp}). These polynomials then constitute the polynomial vector which is a solution and is in the module of solutions over the quotient ring $\Kbar$. In terms of the time-delay operators $x, l, z, n$, the polynomials $p_1$ and $q_1$ are of degree 22, while $q_2$ and $p_3$ are polynomials of degree 23. For a general value of $n$, the solution contains polynomials of maximum degree $8n - 1$ in the time-delay operators. 
\par
Clearly, from the mathematical point of view we have an infinite family of solutions. Note that here we do not lay any claim to any exhaustive listing of solutions. However, we believe this family of solutions is sufficiently rich, because we can form linear combinations of these solutions and then these form a left submodule of the full left-module of solutions. We can also subtract solutions like (\ref{X1'}) from (\ref{X2}) which is also a solution with each component polynomial having less number of terms (or links in the usual terminology) albeit of the same degree. Further, by using cyclic permutations on $(x, y, z)$ and $(l, m, n)$ we can generate more solutions. For example, we may set $a = my$ and $b = zn$ and obtain solutions for the four links, $y, z, m, n$. The linear combinations of these solutions with the previous ones in the variables $x, z, l, n$ are also solutions in the full module pertaining to all six links. Thus we see that a plethora of solutions can be so generated.
\par
From the physical point of view, since we have neglected terms in ${\ddot L}$ and $\Ldot^2$ and higher orders in our calculations, a limit on the degree of the polynomial solutions arises. That is upto certain degree of the polynomials, we can safely assume the commutators to vanish. But as the degree of the polynomials increases it is not possible to neglect these higher order terms and then such a limit becomes important. We now investigate this limit and make a very rough estimate of it. Our LISA model in \cite{DNV08} gives ${\ddot L} \sim 10^{-6}$ m/sec$^2$. The other term $\Ldot^2/c^2 \sim 10^{-15}$ is way too small to matter. So the limit is set by $\ddot L$. From $\ddot L$ we compute the error in $L$, namely, $\Delta L \sim \h \Delta t^2 {\ddot L}$. If we allow the error to be no more than say 10 metres, then we find $\Delta t \sim 4500$ sec. Each time-delay is about $16.7$ secs which gives the number of successive time-delays to be about 270. This is the maximum degree of the polynomials. This means one can go upto $n \lsim 30$. If we set the limit more stringently at $\Delta L \sim 1$ metre, then the highest degree of the polynomial reduces to about 80 which means one can go upto $n = 10$. This means there are a large number of TDI observables available to do the physics. We however point out that when we go to higher degree polynomials, the number of interpolations of data also increases leading to accumulation of numerical  errors. This is a relevant issue whenever high degree polynomials are used.

\section{Conclusion}

In this work we have given a family of approximate solutions for the general model of LISA where the up-down links are unequal and the armlengths are time varying, but with only two arms functional. Such a situation can arise when one of the arms is non-functional or when it is not possible to obtain data from all optical links because high costs etc. We then have only four links to deal with and although the general non-commutative problem of obtaining second generation TDI is extremely difficult, the problem here reduces to a much simpler and tractable one. We have therefore been able to exhibit a family of solutions although not exhaustive, is quite rich and interesting both from the mathematical and physical point of view. Also the mathematical analysis presented here could be extended to other space missions such as ASTROD where a similar situation arises although the physical parameters will be quite different.   

\section{Acknoweledgements}

The authors S. V. Dhurandhar and J-Y Vinet would like to thank the Indo-French Centre for the Promotion of Advanced Research (IFCPAR) project no. 3504-1 under which this work has been carried out.

\end{document}